# A Reassessment of 5f Occupation in Plutonium

JG Tobin
U. Wisconsin-Oshkosh, Oshkosh, WI 54937
Email: tobinj@uwosh.edu

**Abstract**

A detailed spectroscopic simulation of the original Pu $N_{4,5}$ and $O_{4,5}$ X-Ray Absorption Spectroscopy (XAS) has been performed. Additionally, a fundamental flaw in the Electron Energy Loss Spectroscopy (EELS) measurement has been corrected. Thus, the determination of the 5f occupation (n) in elemental Pu has been re-evaluated with the result that n = 5.0 ± 0.1 for αPu and n = 4.9 ± 0.2 for δPu. These values are significantly lower than the value of ~5½ that was propagated earlier.

**I        Introduction**

Whether loved, hated, feared, reviled or condemned, plutonium is important, because of its nuclear energetic properties, i.e. fission. [1] It is also very unusual and quite interesting as a material. It is chemically toxic, highly radioactive, and pyrolytic, with an exasperating tendency to catch fire spontaneously. Unlike normal metals, its most dense phase (α) is not face-centered-cubic (fcc) or hexagonal close packed: it is lower symmetry monoclinic. The fcc phase (δ) is substantially less dense. It has six solid phases elementally. It also shares with water another rare trait: it expands with freezing. [1,2,3] This physical complexity is mirrored in its electronic structure and spectroscopy, both for the actinides in general and Pu in particular. [1-20]

The Pu 5f occupation (n) is a crucially important parameter in the understanding of 5f electronic structure. While it is clear that n ~ 5 must be true [4-6], the exact value could

# A Reassessment of 5f Occupation in Plutonium

potentially eliminate many of the candidate models for the explanation of Pu electronic structure. [7-11]. A value of n near 5½ is very different than a value of n near 5.0, as will be discussed below. The original Pu $N_{4,5}$ and $O_{4,5}$ XAS [2,3,12-14] (Figure 1) has been quantitatively reevaluated, including using the FEFF spectral simulation program, a Green's function based, multiple scattering code. [21-31] The details of the FEFF simulation is discussed below. The result of this analysis, including comparison to U [32] is that $n_{\alpha Pu}$ = 5.0 ± 0.1 and $n_{\delta Pu}$ = 4.9 ± 0.2. Additionally, a long-standing problem with the EELS measurements has been resolved.

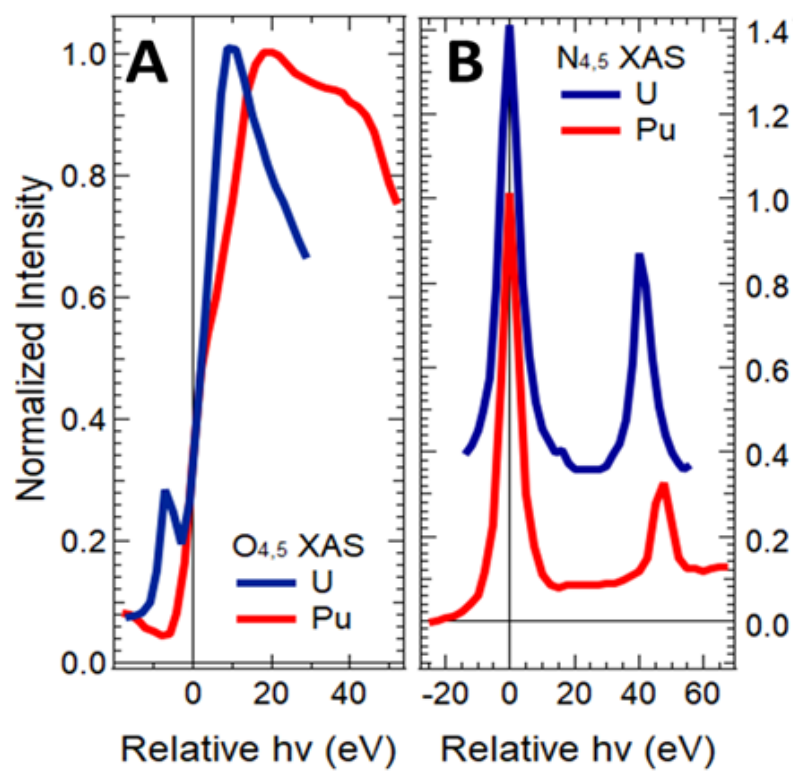

Figure 1  Shown here are the (A) $O_{4,5}$ and (B) $N_{4,5}$ XAS of elemental uranium (α-U) and plutonium(α-Pu). The features have been vertically normalized to unity and the spectra shifted horizontally. For the $O_{4,5}$, the spectra were aligned at the edge jump and lowest hv. The $N_{4,5}$ spectra were aligned and normalized at the $N_5$ maximum. The relative energy scales are preserved. The Pu (U) spectra are from Ref [2,3, 12-14] ([32]).

Owing to the recent renewed interest in the determination of 5f occupation in actinide materials [15,16] a quantitative evaluation of the 2$^{nd}$ Derivative mode in Electron Energy Loss Spectroscopy (EELS) was undertaken. [17] It was found that, while EELS does converge to an X-ray Absorption Spectroscopy (XAS)-like limit (electron dipole transition) at high energies, [18] there are significant issues with the 2$^{nd}$ Derivative Mode, utilized in the work by Moore et al. [19]

# A Reassessment of 5f Occupation in Plutonium

The 2nd Derivative Analysis is the underpinning of the proposed result that n = 5.4. Earlier studies utilizing a combination of both XAS and EELS analysis done in the non-derivative mode [13,14,20] did not exhibit the same problems as the EELS studies performed with EELS only in the 2nd Derivative Mode. Herein, it will be shown that a correction can be applied to the 2nd Derivative spectra that results in $n_{\delta pu}$ ~ 5.0 also.

In the next section, the computational and experimental specifics will be considered, followed by sections with a discussion of the results and then conclusions.

II      **Computational and Experimental**

FEFF is a data analysis platform used in x-ray absorption spectroscopy and associated techniques. It includes self-consistent real space multiple-scattering with simultaneous calculations of x-ray absorption spectra and the electronic structure, with special emphasis on the extended x-ray absorption fine structure (EXAFS). [21-24] The name FEFF originated with $f_{eff}$ or f-effective, a term from x-ray scattering theory. FEFF's beginnings trace directly to the requirement to quantify the elastic electron scattering events in the EXAFS final state, giving it a sensitivity to interatomic distances. The JFEFF version of FEFF 9.9.1 was used for these calculations. FEFF is a Green's function based approach, having been shown to be an accessible, robust, and very effective tool for the analysis of x-ray measurements. In the case of actinides (5f materials), the application of FEFF has been historically long standing and efficacious, but with some limitations in terms of the treatment of the 5f states. [25-31] In the FEFF calculations, the δ-Pu was modelled with a 79-atom cluster with face-centered cubic order and a nearest neighbor distance of 3.28Å.

# A Reassessment of 5f Occupation in Plutonium

Table 1

The δ-Pu $O_{4,5}$ XAS FEFF parameters were as follows:

| | | |
|---|---|---|
| XANES 3,0.05, 0.3; FMS 3.0; | Exchange off; | CHWIDTH 2; |
| SCF 30,0,100,0.2,1; | Core Hole on, No; | Absolute Cross Section; |
| UNFREEZE; | SO2 off; | Debye 190,315; |

Additional detail is provided in the references [25-31] and the results section below. The XAS measurements were carried out at the Advanced Light Source on Beamline 7.0. as described in detail elsewhere. [2,3] In general, experimentally, alpha phase samples are pure Pu, delta phase samples are mainly Pu with small amounts of Ga to stabilize the delta phase at room temperature and alpha-prime will be alpha phase with a small amount of Ga. The simulations are pure Pu.

Next, the spectroscopic results and simulations will be addressed.

### III    Results and Discussion

The most important spectroscopic data for Pu come from the X-Ray Absorption Spectroscopy (XAS) of the $O_{4,5}$ ($5d_{5/2}$ and $5d_{3/2}$, 5d→5f, hv ~ 100 eV) and $N_{4,5}$ ($4d_{5/2}$ and $4d_{3/2}$, 4d→5f, hv ~ 800 eV). In Figure 1, the elemental Pu measurements are compared to those of elemental U. Two effects are immediately obvious. In the $O_{4,5}$ spectra, there is a small pre-peak in the U but not the Pu. In the $N_{4,5}$ spectra (Relative hv ~ 40 eV), the $N_4$ peak in Pu is significantly reduced compared to that of the U. These are the spectroscopic fingerprints of n = 5 in Pu. The reduction of the $N_4$ peak in Pu is driven by angular momentum coupling, as predicted by the Intermediate Coupling Model (ICM). [13,14,33] The relative intensity of the $N_4$ peak is best discussed in terms of the Branching Ratio: BR = $I_{N5}/(I_{N5} + I_{N4})$. The $N_{4,5}$ spectra are a

# A Reassessment of 5f Occupation in Plutonium

perfect case for the use of the BR: the peaks are far enough apart to be separately analyzed and yet still in the same spectrum, thus retaining self-normalization. Using the average of the XAS and EELS values in the original study, for α-Pu, $BR_{\alpha Pu}$ = 0.82; for α-U, $Br_{\alpha U}$ = 0.68. [13] These values are in excellent agreement with our understanding from the Intermediate Coupling Model (ICM), with $BR_{n=5, localized}$ = 0.817 [13], and recent work on 5f delocalization [34], where and $BR_{n=3, delocalized} \approx BR_{n=2, localized}$ = 0.68. The loss of the $O_{4,5}$ pre-peak, which is normally observed in many Rare Earths and light Actinides, is caused by the filling of the $5f_{5/2}$ manifold in the final state: $4d^{10}5f_{5/2}^5$ + hv → $4d^9 5f_{5/2}^6$. As originally explained by Dehmer at al [35], the pre-peak effect in the Rare Earths and Actinides is driven by angular momentum coupling between the core hole and the f-state hole. [12] The filled $5f_{5/2}^6$ manifold quenches the final state coupling effect.

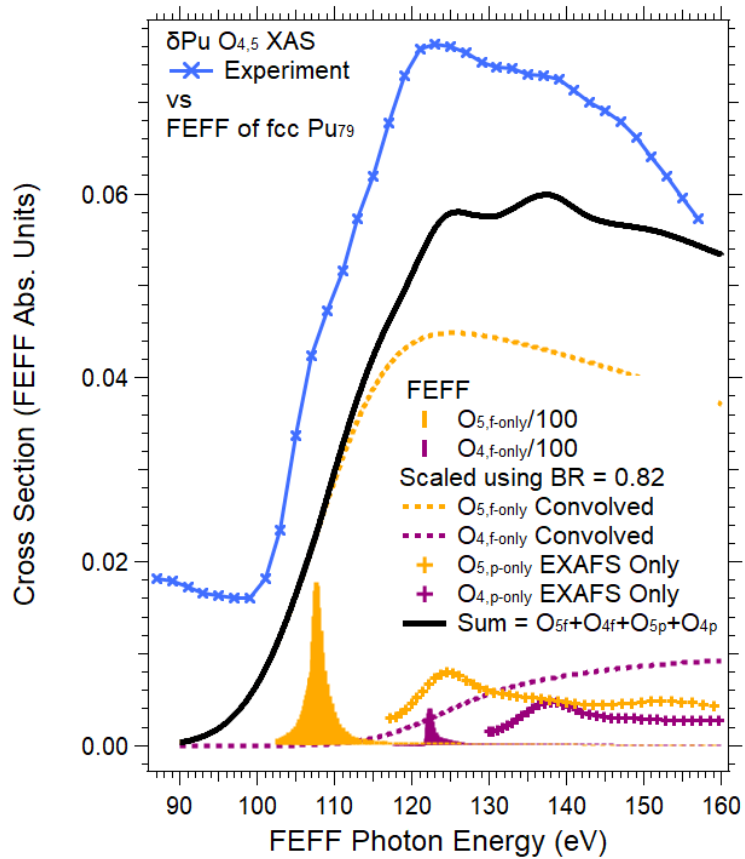

Figure 2    Shown here is a comparison of the Pu $O_{4,5}$ XAS experiment and the FEFF simulation using the 79-atom cluster. There is no core hole coupling and no pre-peak. The d→f transitions have been convolved with an asymmetric gaussian, to simulate the Fano line shape. Some of the peak structures (~125 eV, ~140 eV and ~ 150 eV) are caused by EXAFS peaks, as can be seen in the d→p channels. The asymmetric gaussian = fexpgauss (hv – 110, 0.01, 70), following Ref [36]. The p contributions are EXAFS only. The experimental spectrum has been shifted horizontally and shifted and scaled vertically.

# A Reassessment of 5f Occupation in Plutonium

In Figure 2, the observed $O_{4,5}$ XAS for δ-Pu has been simulated using FEFF and convolution with an asymmetric gaussian [36], with a high level of agreement. (δ-Pu has the great advantage of simplicity and order over α-Pu: while the fcc δ-Pu has a single site, the monoclinic α-Pu has 16 atoms in the unit cell, with eight unique sites. [14] No gallium impurities are used in FEFF.) There are four fundamental peaks in the spectrum: d→f and d→p for each of the two edges, $O_5$ ($5d_{5/2}$) and $O_4$ ($5d_{3/2}$). The broadening of the Pu $O_{4,5}$ is caused by the very strong Fano line-shape previously observed in the Resonant Photoelectron Spectroscopy (ResPES) of Pu. [2,3] However, the Fano line-shape can only be applied to the d→f channels. If the Fano line-shape is used with the d→p transitions, the fine structure (hv ~ 125 eV, ~ 140 eV and ~ 150 eV) is lost. To get this level of agreement, it is also necessary to properly scale the 5f intensities: from BR = 0.82, I4/I5 = 0.22 was applied to the d→f white-line peak heights. However, the goal is to independently determine the BR and for that, one must turn to the α-Pu $N_{4,5}$ XAS.

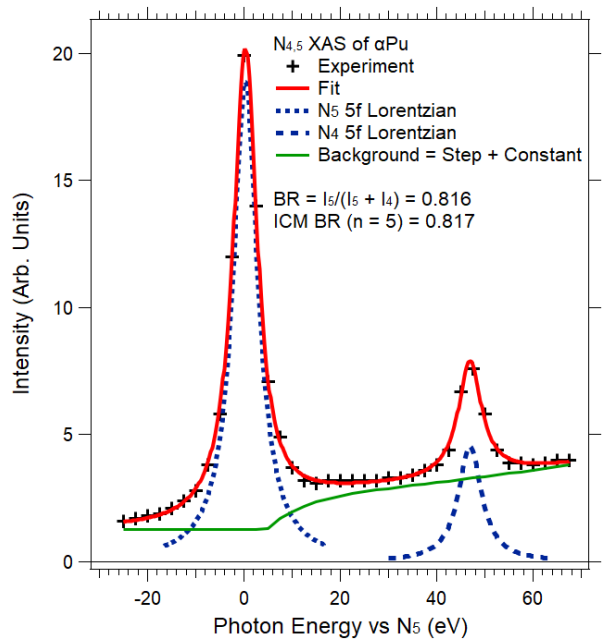

*Figure 3    The peak fitting of the αPu $N_{4,5}$ XAS is presented here. Lorentzian peak-shapes have been used for the $N_4$ and $N_5$ 5f white-lines and a step function with a constant is used to model the $N_5$ 7p/EXAFS contribution. The step function was an exponential convolved with an exponential. There is a remarkable level of agreement with $BR_{XAS}$ = 0.816 and $BR_{ICM,n=5}$ = 0.817. The spectrum was taken from Ref [13,14].*

# A Reassessment of 5f Occupation in Plutonium

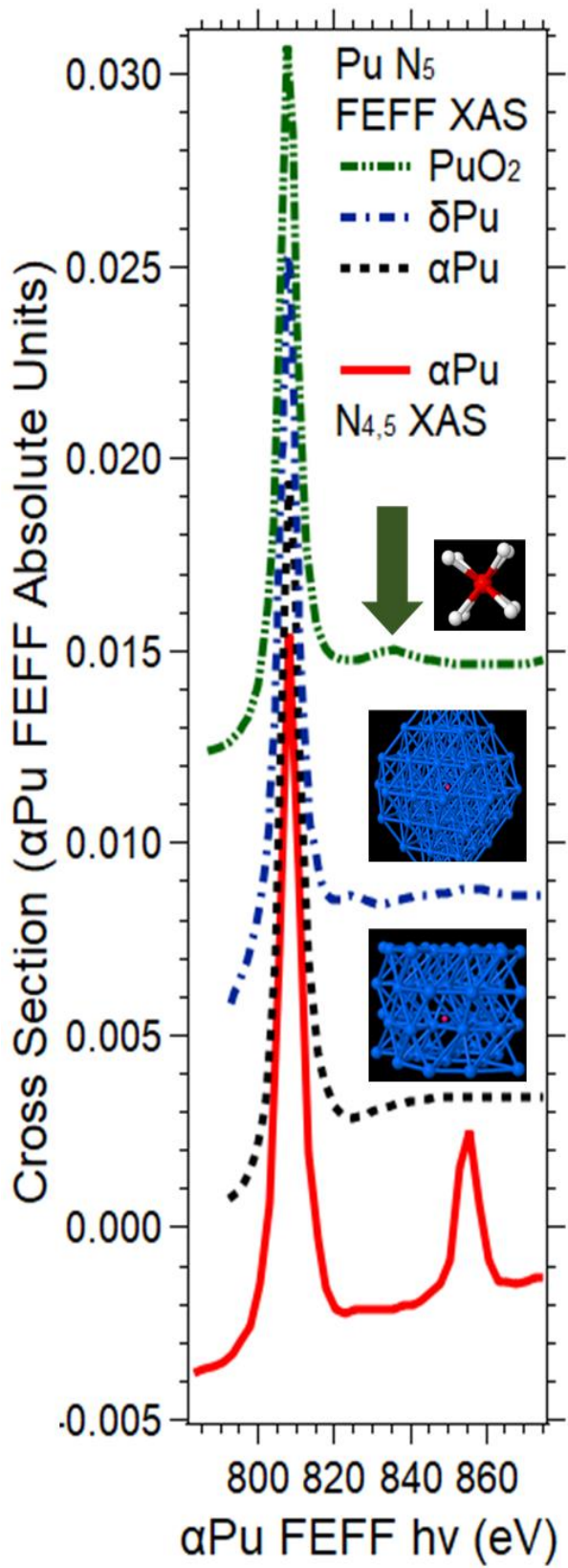

The peak fitting of the αPu $N_{4,5}$ XAS is plotted in Figure 3. Here, three functions were used to perform the fitting: a Lorentzian for the $N_5$ 5f peak, a step with a constant for the $N_5$ 7p/EXAFS feature and another Lorentzian for the $N_4$ 5f peak. The level of agreement very good, possibly too good: $BR_{XAS}$ = 0.816 and $BR_{ICM,n=5}$ = 0.817. If the fitting functions are changed slightly, using a linear function for the background along with two Lorentzians, similar but not identical results are obtained: $BR_{XAS}$ = 0.805. While this result is an indication of the robust nature of the BR determination, it begs the question: which 7p/EXAFS or background function is more appropriate? To resolve this issue, it is necessary to resort to FEFF simulations again.

Figure 4  FEFF simulations for the Pu $N_5$ XAS of a series of Pu materials are illustrated here. The corresponding model structures are included as insets: The Pu emitter is red; the oxygen atoms are white; surrounding Pu atoms are blue. From the top: $PuO_2$, modeled with a $PuO_8$ cluster with local octahedral/cubic symmetry [16]; δPu, modeled with the 79-atom fcc cluster described earlier; αPu, modeled with a 54-atom cluster from Ref [37]. The αPu emission site was arbitrarily chosen to be near the center of the cluster. The experimental αPu $N_{4,5}$ spectrum is bottommost, for comparison. Other than the FEFF of αPu, the spectra have been shifted horizontally and scaled and shifted vertically.

JESRP                              7                         11 March 2024

# A Reassessment of 5f Occupation in Plutonium

The FEFF simulations can be seen in Figure 4. Both the $PuO_8$ model for fluorite $PuO_2$ and the $Pu_{79}$ fcc cluster for δPu have the great advantage of the simplicity of a single site material. It must be admitted that simulations for αPu [37] do not have that simplicity, because the monoclinic unit cell has 16 atoms with 8 distinct sites. Nevertheless, it is useful to consider the FEFF results for the arbitrarily chosen emission site in Figure 4. Clearly, the 7p/EXAFS feature in the αPu FEFF is step-like. On the other hand, the $PuO_8$ spectrum has a fairly sharp and intense peak, as accentuated with the green arrow. This 7p/EXAFS peak can be seen all of the closely related $ThO_2$, $UO_2$ and $PuO_2$ compounds. [16, 31, 38-40] The δPu case is somewhere in the middle, with either a less distinct peak or a less constant step. In any case, the FEFF results indicate that the step function is more appropriate for αPu.

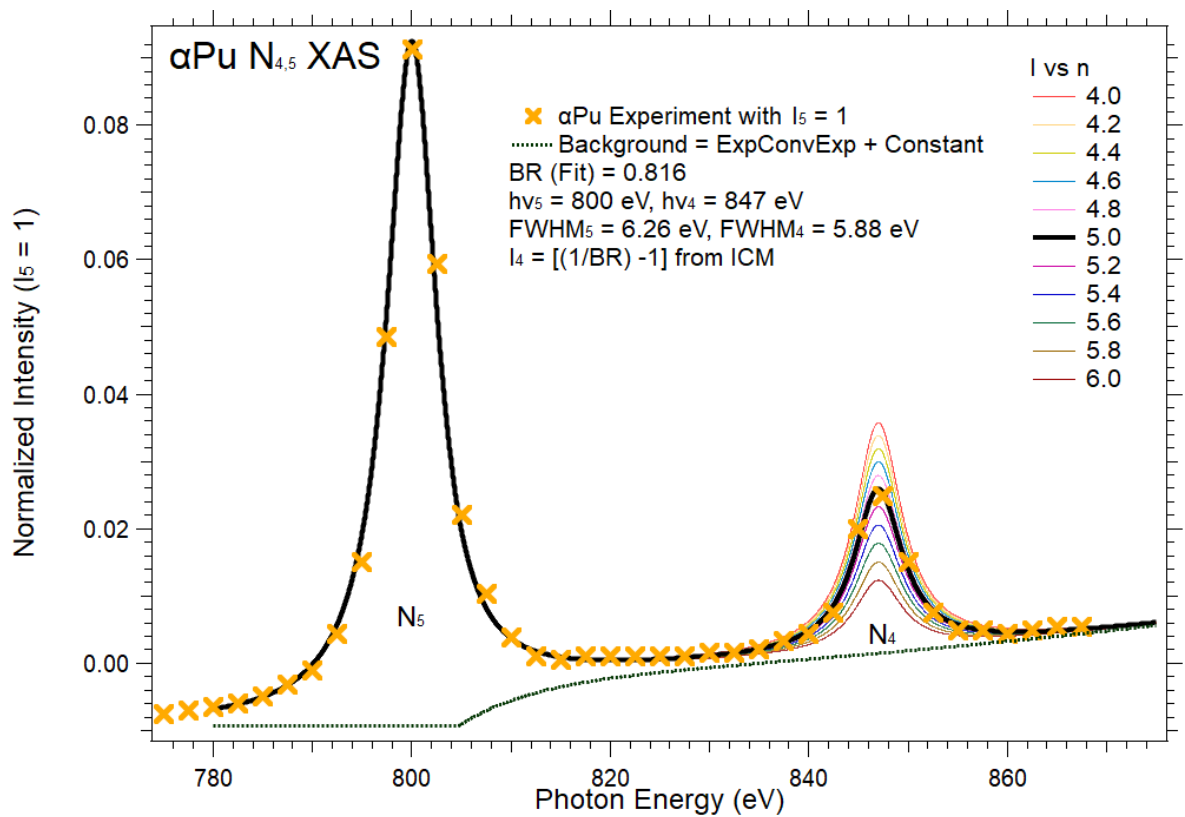

Figure 5    Quantifying the n determination from the αPu $N_{4,5}$ XAS. Here, the $N_4$ integrated peak intensity is scaled with n, using $I_{N4}/I_{N5} = ((1/BR) – 1)$ and linear interpolation. As before, the $N_5$ integrated peak intensity is fixed at unity. From this simulation, it is found that $\Delta n = \pm 0.1$. See text for further discussion.

# A Reassessment of 5f Occupation in Plutonium

At this point, it is useful to quantify the n determination for αPu. As shown in Figure 5, the $N_4$ intensity can be scaled with n. Keeping the $N_5$ integrated peak intensity at unity, the $N_4$ integrated peak intensity will follow the relation $(I_{N4})_{IN5 = 1} = ((1/BR) -1)$, using the ICM values and linearly interpolating using BR = 1 - (14/15)($N_{5/2}$/N), where $N_{5/2}$ is the number of 5f holes in the j = 5/2 manifold and $N$ is the total number of 5f holes. [13,14,34] From inspection of the plot, it can be seen that n > 5.2 and n < 4.8 are not consistent with the experimental data. Thus, a conservative error estimate would be Δn = ± 0.1 and ΔBR = ± 0.01, which is also consistent with the estimate of $BR_{XAS}$ = 0.805, obtained using a linear background. This results in the determination that $n_{αPu}$ = 5.0 ± 0.1. The question of $n_{δPu}$ is addressed next.

In general, the Pu $N_{4,5}$ XAS of are almost identical for multiple phases, ages, and doping of elemental Pu. [13, 14, 19, 41] In the original studies done with XAS and EELS, the BR were essentially the same, especially when considered within the limits of the error estimate of ΔBR = ± 0.01: $BR_{αXAS}$ = 0.813 and $BR_{αEELS}$ = 0.826. Another example of this is the direct comparison of spectra shown in Figure 6.

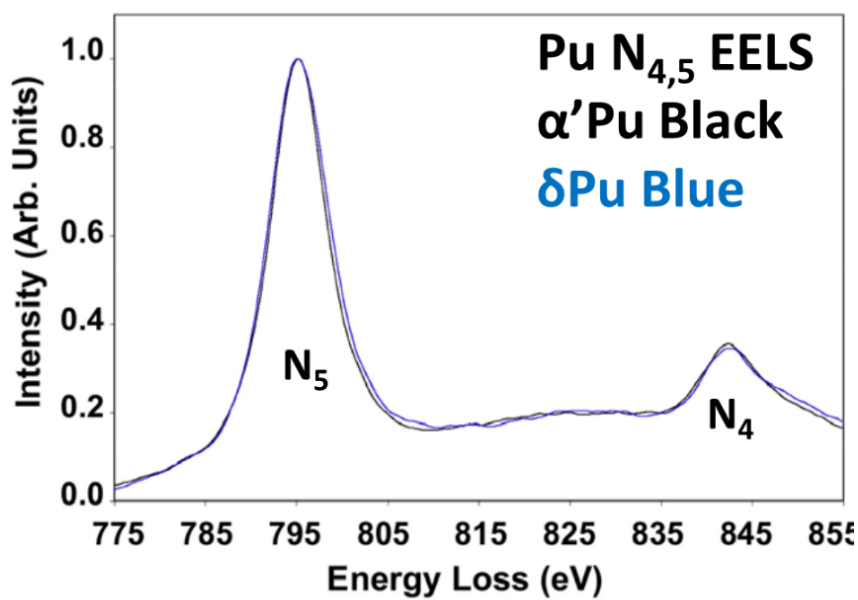

*Figure 6. The EELS $N_{4,5}$ data for the 4d to 5f transition in α'Pu and δPu, taken from Ref [41] α'Pu = black/dark line. δPu = blue/light line. Here, α'Pu denotes a small Ga impurity in the α lattice.*

# A Reassessment of 5f Occupation in Plutonium

Later studies of the BR in Pu materials switched over to EELS only and the use of the 2$^{nd}$ Derivative Spectrum in EELS. It has been demonstrated recently that the 2$^{nd}$ Derivative method is fundamentally flawed. [17] Fortunately, it is possible to correct for the error in the 2$^{nd}$ Derivative method. That will be discussed shortly.

However, before going on to that issue, it is useful to consider the results of fitting a zero derivative (non-derivative) EELS spectrum of δPu. [19] This is essentially the equivalent of an XAS spectrum. The results are plotted in Figure 7: $BR_{\delta EELS}$ = 0.81 ± 0.02, nearly the same as for αPu XAS, $BR_{\alpha XAS}$ = 0.816. Several variations on the fitting were run, with results generally consistent with those shown in Figure 7. For example, using Voigt functions instead of Lorentzians also results in BR = 0.81. However, varying the background/EXAFS modelling leads to degradation in the quality of the fit. The superiority of the linear function and gaussian peak combination is consistent with the simulations in Figure 4, where the EXAFS feature structure is between that in $PuO_2$ and αPu. The error estimate of ΔBR = ±0.02 comes directly from the application of the fit uncertainties, i.e. the ± integrated area values for the two Lorentzians.

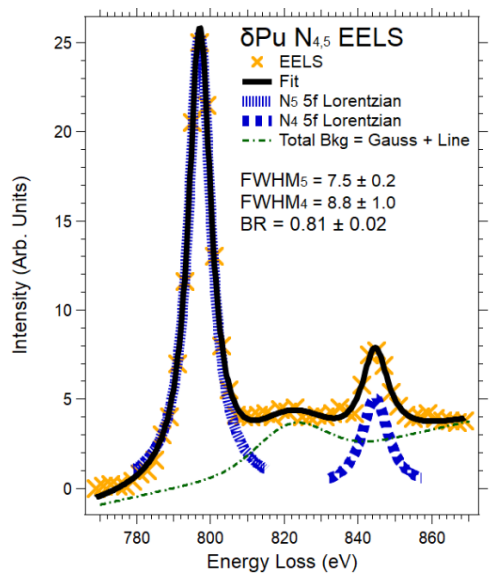

*Figure 7    The results of peak fitting the zero derivative (non-derivative) $N_{4,5}$ EELS of a new δ-Pu sample are shown here. Lorentzian functions were used for the 5f peaks, a gaussian for the $N_5$ EXAFS peak and a linear function for the background. The spectrum was taken from Ref [19]. FWHM is the full-width at half maximum of the Lorentzian functions for the $N_4$ and $N_5$ 5f white-lines, including the errors.*

# A Reassessment of 5f Occupation in Plutonium

Returning to the issue of the 2nd derivative analysis, the problem is not with the non-derivative peak analysis in EELS but rather the changeover to a complete reliance upon the 2nd derivative analysis. As can be seen in Panel A of Figure 8, the XAS and EELS spectra of the αPu are very similar, with the only exception being the enhancement of the EXAFS peak near an energy of 20eV in the EELS spectrum. This enhancement of the EXAFS peak in EELS is not understood, but is commonly observed. [19] An example of the problems encountered is shown in Figure 8, Panel B. Here, using normalized Lorentzian peak-shapes, the peak ratio is 1 ($I_{0,4}/I_{0,5}$ = 1, where $I_0$ is the peak intensity of the non-derivative peaks) and the Branching Ratio (BR) is ½. ($BR_0 = I_{0,5}/(I_{0,5} + I_{0,4})$) On the other hand, the 2nd Derivative results are as follows: $I_{2,4}/I_{2,5}$ = 0.64, where $I_2$ is the peak intensity of the non-derivative peaks, and $BR_2 = I_{2,5}/(I_{2,5} + I_{2,4})$ = 0.6. (As will be discussed below, the 0.64 comes from the 2nd order dependence upon the 2nd Derivative peak widths: $(4/5)^2$ = 0.64.)

While the 2nd Derivative mode is a powerful approach for the removal of background peaks (low frequency noise), there are several problems that limit its utility for quantitative analysis. The derivation of these results is provided in Reference [17]. Below is a summary of these results.

1. There is no fixed relationship between non-derivative peak intensities and 2nd derivative peak intensities. Changing peak-shapes changes the ratio.
2. Assuming that peak-shapes remain constant, which may not be justified, there is a second order dependence of the ratio upon the peak width of the 2nd Derivative peak. An example of this is shown in Figure 8, Panel B.
3. The second order dependence upon the peak width manifests itself in two very distinct ways: with random errors and systematic errors

# A Reassessment of 5f Occupation in Plutonium

a. The random error from high frequency noise places a limit on the number of significant digits in the BR result.

b. The systematic error has shown up in the BR predictions for the localized systems Pu, $PuO_2$ and $UO_2$ [42]. (The delocalized U seems to have other broadening effects that remove the problem.) Each of these cases is discussed below.

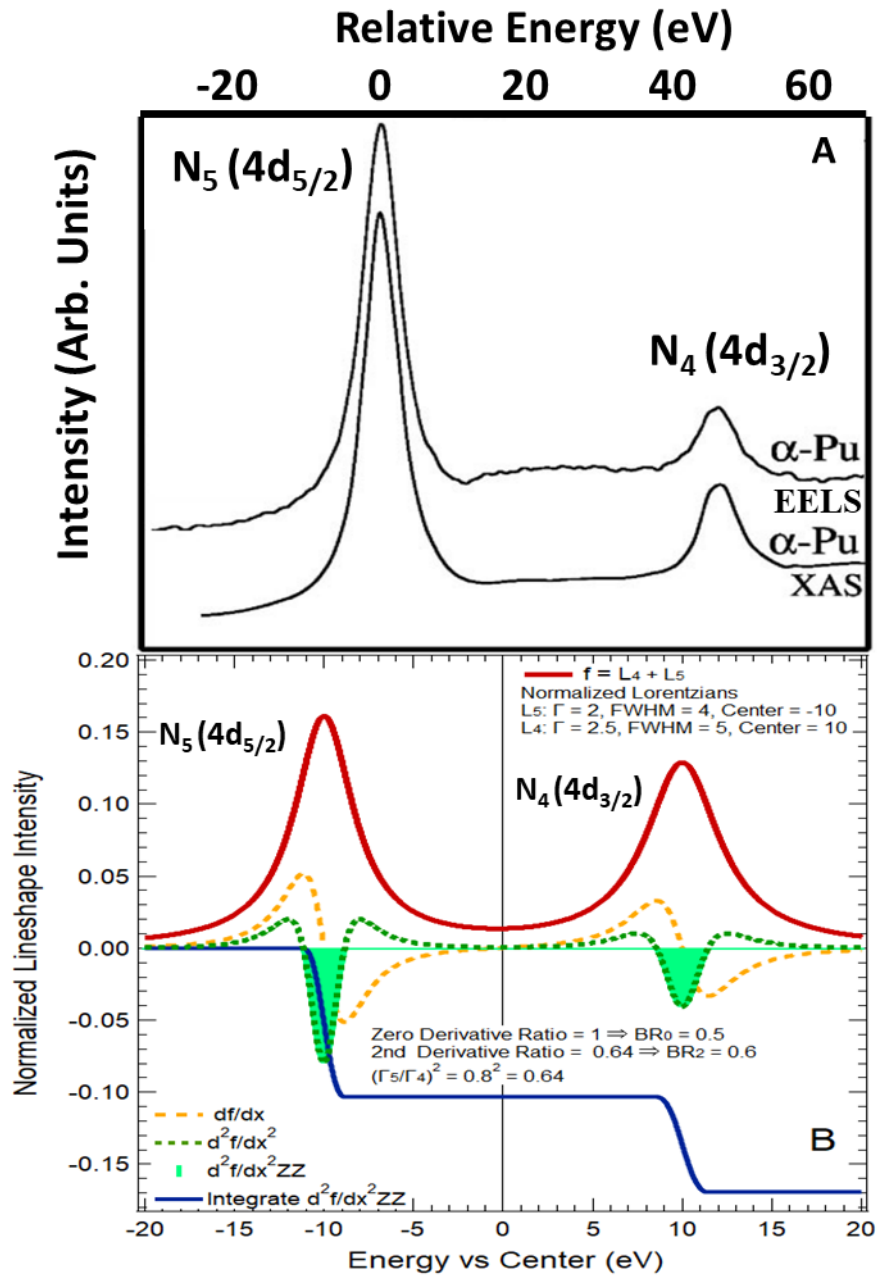

Figure 8

# A Reassessment of 5f Occupation in Plutonium

*Figure 8*

*Panel A: Shown here is a comparison of the non-derivative spectra for the $N_{4,5}$ XAS and EELS of the of alpha-Pu. This data is taken from Ref. [13,14]. Note the strong similarity of the XAS and EELS Spectra. The primary beam of the EELS experiment was near 300,000 eV. The X-rays used were in the range of 770 to 870 eV. In the original papers, a small artifact was removed from the high energy side of $N_4$ XAS peak, possibly caused by absorption in the beamline optics.*

*Panel B: A series of simulated spectra are shown here: the non-derivative, first derivative and second derivative. The normalized Lorentzian non-derivative peaks each have an area of unity and a ratio of unity. The $N_4$ peak is wider, with a Full Width at Half Max (FWHM) of 5 eV ($\Gamma$ = ½ FWHM = 2.5 eV), while the $N_5$ has a FWHM = 4 and $\Gamma$ = 2 eV. The 2$^{nd}$ Derivative peaks, defined as being from zero to zero as shown, deviate from unity in both the raw values and in the ratio. See the text for further discussion.*

**Random Error**

High frequency noise such as the Poisson Statistics of spectroscopic investigations place fundamental limitations upon the number of significant digits in the result. In Figure 9, the impact of high frequency noise upon a smooth curve is illustrated. Here, artificial noise is added to a noiseless spectrum. The value of the noise was chosen to roughly match the magnitude and frequency of the noise in experimental spectra. [17] The sinusoidal magnitude is 0.0002 and the sinusoidal frequency is such that there is a 2π change in the argument every 5 eV. The variation in ZZ, the base-widths of the 2$^{nd}$ Derivative Peaks, drives an uncertainty in the peak intensities. Following the analysis in Ref. [17] and using BR = 0.68 and Area$_{N4}$/Area$_{N5}$ = 0.47, the error estimated below is obtained.

$\Delta BR = \pm \{BR\}(Area_{N4}/Area_{N5})\{(0{\cdot}03)^2 + (0{\cdot}06)^2\}^{½} = \pm\ 0.02$

# A Reassessment of 5f Occupation in Plutonium

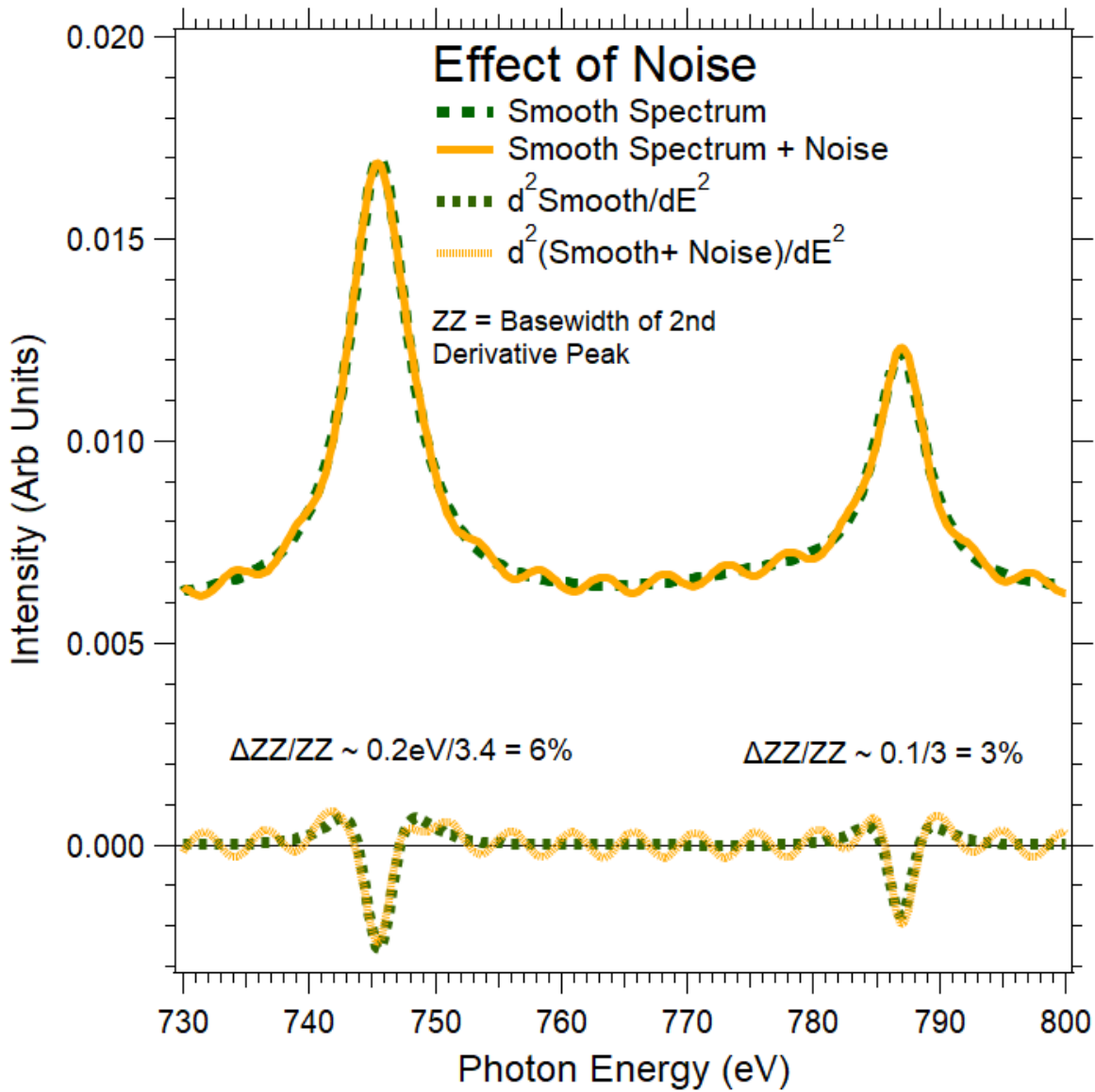

*Figure 9                A comparison of the second derivative peaks for smooth curve and the smooth curve plus high frequency noise. Note the change in ZZ. See the text for further discussion. Similar to Figure 11 of Ref. [17].  $N_5$ is near hν = 745 eV and $N_4$ is near hν = 787 eV. See Ref [17] for further detail.*

One salient result would thus be the following: None of the Branching Ratios in the paper by Moore et al would be valid beyond ±0.02, rendering many of the results indistinguishable.  For example, the metallic Pu results in Table 2 would all be statistically the same, with the $PuO_2$ only ever so slightly different.

# A Reassessment of 5f Occupation in Plutonium

Table 2: The impact of the error limit of ΔBR = ±0.02 upon the original analysis.

| Material | 2nd Deriv 5f count | 2nd Deriv Branching ratio |
|---|---|---|
| alpha-Pu new | 5.4 | 0.843±0.02 |
| alpha'-Pu new | 5.4 | 0.842±0.02 |
| delta-Pu new | 5.4 | 0.847±0.02 |
| delta-Pu aged | 5.4 | 0.856±0.02 |
| $PuO_2$ | 5.4 | 0.874±0.02 |

**Systematic Error**

However, there is another, more important problem lurking in this data: a systematic shift caused by peak width variation. This systematic shift is what forces the 2nd Derivative 5f counts in Table 1 to be too high.

Consider the four cases shown in Table 3. For comparison, the XAS result for αPu is also utilized, to avoid the problem with the enhanced EXAFS in the EELS.

Table 3

| Actinide 5f Material | BR XAS and non-derivative EELS | BR 2nd Derivative EELS | Corrected BR From 2nd Derivative EELS | n ± Δn From XAS/EELS and ICM** |
|---|---|---|---|---|
| αPu | **0.816 ± 0.01*** | 0.843 | 0.816 | **5.0 ± 0.1*** |
| δPu | **0.81 ± 0.02*** | 0.847 | 0.821 | **4.9 ± 0.2*** |
| $PuO_2$ | **0.85 ± 0.05**[16] | 0.874 | 0.851 | **5.3 ± 0.6*** |
| $UO_2$ | **0.68 ± 0.02**[42] | 0.729 | 0.690 | **2.0 ± 0.4*** |

*This study
**Interpolated using BR (n = 6) = 0.918, BR (n = 5) = 0.817, BR (n = 4) = 0.760, BR (n = 3) = 0.723, BR (n = 2) = 0.680, BR (n = 1) = 0.634. From Ref {13}. Error estimates in n were based the average of the two values, rounding downwards.

# A Reassessment of 5f Occupation in Plutonium

There is anecdotal evidence that for EELS, the width of the $N_5$ EELS peak is nearly the same but slightly smaller than the width of the $N_4$ EELS peak. An example of this can be seen in Figure 7. Assuming that for these localized systems, there is a 10% reduction in the width of the $N_5$ versus the $N_4$, the result in the last column is obtained. (The correction is made by extracting the peak ratio ($I_4/I_5$) from the 2$^{nd}$ Derivative BR, multiplying the peak ratio ($I_4/I_5$) by $(1.1)^2 = 1.21$, and calculating the corrected BR from the adjusted peak ratio.) With the correction for different widths, the EELS results return to agreement with the XAS/EELS and the 5f populations are reduced, to $n^{5f}(Pu) \sim 5.0$ and $n^{5f}(U) \sim 2.0$, as is commonly accepted. [12-14, 16, 17, 20, 42]

<u>The 2$^{nd}$ Derivative Mode in EELS should not be used in quantitative BR analyses without correction for these effects.</u>

## IV      Conclusion and Summary

Table 4 Elemental Plutonium 5f occupations

| Year | Method | αPu n | δPu n |
|---|---|---|---|
| **2004/2005** | **XAS & EELS [13,14]** | **5** | **5** |
| 2005 | LDA + U [43] | 6 | 6 |
| 2006 | 2$^{nd}$ Deriv EELS [19] | 5.4 uncorrected | 5.4 uncorrected |
| 2006 | LDA+U/DMFT [9] | --- | ~5½ |
| 2007 | DMFT Theory [8] | --- | 5.2 |
| 2010 | AIM [45] | 5.2 | 5.2 |
| 2012 | RXES [46] | 5.2 | 5.3 |
| 2014 | RXES [47] | 5.2 | 5.4 |
| **2015** | **DMFT Theory [11]** | **---** | **5.04** |
| **This work** | **This work** | **5.0 ± 0.1** | **4.9 ± 0.2** |

It is useful to consider the Pu 5f occupation from an historical perspective. (See Table 4.) Circa 2004/2005 the first XAS/EELS results came out [13,14], with strong indications that n ~ 5 for both alpha and delta Pu. Shortly thereafter, Shorikov and coworkers [43] published an LDA + U (Local Density Approximation + U) calculation that suggested that n ~ 6, driven by the desire

# A Reassessment of 5f Occupation in Plutonium

to obtain magnetic cancellation in the simulation of the Pu electronic structure. [44] About the same time, Moore utilized the 2nd derivative method in EELS to obtain the erroneous conclusion that n = 5.4. This seems to have sent the field in the wrong direction, with a number of studies tilted towards the mid 5+ range. [8,9,45-47] Pourovski et al [9], using an LDA+U method coupled to DMFT (Dynamical Mean Field Theory) produced a result that suggested that n ~ 5½ and Booth and coworkers, utilizing Resonant X-ray Emission Scattering (RXES) [46,47], promulgated 0.52 < n $\leq$ 5.4.  Even Shim, Haule & Kotliar [8], working with DMFT, and van der Laan and Taguchi [45], taking an Anderson Impurity Model (AIM) approach, came up with n = 5.2. However, finally in 2015, Janoschek and colleagues obtained the result that n = 5.04 for delta Pu. This is the only value in quantitative agreement with the results of this study.

Finally, some key points to summarize:

1. $n_{\alpha Pu}$ = 5.0 ± 0.1
2. $n_{\delta Pu}$ = 4.9 ± 0.2.
3. The 2nd Derivative Mode in EELS should not be used in quantitative BR analyses without correction for skewing effects.

ACKNOWLEDGEMENTS: The author wishes to thank the University of Wisconsin-Oshkosh and its faculty and staff for their ongoing support and encouragement.  It is a gem set beside the Fox River and Lake Winnebago.  The ALS is supported by the Director of the Office of Science, OBES of the U.S. Department of Energy at LBNL under Contract No. DE-AC02-05CH11231

DATA AVAILABILITY: The data that support the findings of this study are available from the corresponding author upon reasonable request.

AUTHORS DECLARATION: The author has no conflicts to disclose.

# A Reassessment of 5f Occupation in Plutonium